\documentclass[epj]{svjour}
\usepackage{graphics}
\usepackage{graphicx}
\usepackage{amsfonts}
\usepackage{amsmath}
\usepackage{amssymb}
\usepackage{hyperref}
\begin{document}
\title{A conservative energy-momentum tensor in the $f(R,T)$ gravity and its implications for the phenomenology of neutron stars}

\author{S.I. dos Santos Jr.$^1$, G.A. Carvalho$^{1,2,3}$, P.H.R.S Moraes$^1$, C.H. Lenzi$^1$ and M.Malheiro$^1$}% etc
%\thanks{\emph{Present address:} samuel.isidoro.santos@gmail.com.}%                     % Do not remove
\institute{Departamento de F\'isica, Instituto Tecnol\'ogico de Aeron\'autica, S\~ao Jos\'e dos Campos, SP, 12228-900, Brazil
\and
Dipartimento di Fisica and ICRA, 
Sapienza Universit\`a di Roma, 
P.le Aldo Moro 5, 
I--00185 Rome, 
Italy.
\and          
ICRANet,
P.zza della Repubblica 10, 
I--65122 Pescara, 
Italy.}

\date{Received: date / Revised version: date}
\abstract{
The solutions for the Tolmann-Oppenheimer-Volkoff (TOV) equation bring valuable informations about the macroscopical features of compact astrophysical objects as neutron stars. They are sensitive to both the equation of state considered for nuclear matter and the background gravitational theory. In this work we construct the TOV equation for a conservative version of the $f(R,T)$ gravity. While the non-vanishing of the covariant derivative of the $f(R,T)$ energy-momentum tensor yields, in a cosmological perspective, the prediction of creation of matter throughout the universe evolution as shown by T. Harko, in the analysis of the hydrostatic equilibrium of compact astrophysical objects, this property still lacks a convincing physical explanation. The imposition of $\nabla^{\mu}T_{\mu\nu}=0$ demands a particular form for the function $h(T)$ in $f(R,T)=R+h(T)$, which is here derived. Therefore, the choice of a specific equation of state for the  star matter demands a unique form of $h(T)$, manifesting a strong connection between conserved $f(R,T)$ gravity and the star matter constitution. We construct and solve the TOV equation for the general equation of state for $p=k\rho^{\Gamma}$, with $k$ being the EoS parameter, $\rho$ {\it the energy density} and $\Gamma$ is the adiabatic index. We also derive the macroscopical properties of neutron stars ($\Gamma=5/3$) within this approach.
\PACS{
      {PACS-key}{alternative, gravity, theory}   \and
      {PACS-key}{TOV}
     } % end of PACS codes
} %end of abstract
\maketitle
\section{Introduction}
\label{intro}
In the last years, alternative theories of gravity have assumed an important role in the attempts for explaining or evading some shortcomings one faces when considers General Relativity (standard gravity) as the underlying gravitational theory. As some examples of those shortcomings, one could quote the dark energy problem \cite{weinberg/1989,hinshaw/2013} and the observation of massive pulsars \cite{demorest/2010,antoniadis/2013} and white dwarfs \cite{howell/2006,scalzo/2010,kepler/2007}. Today, the most popular of the alternative gravity theories is the $f(R)$ theory \cite{nojiri/2011,sotiriou/2010,de_felice/2010}, which takes a general function of the Ricci scalar $R$ in the gravitational action as its starting point. As it is expected, the presence of general terms in $R$ in the action yields extra terms in the field equations of the theory, and those, in a cosmological aspect, can explain the present cosmic acceleration \cite{riess/1998,perlmutter/1999} with no need for dark energy \cite{amendola/2007,song/2007}. Such extra terms can also elevate the maximum mass expected for neutron stars \cite{capozziello/2016,astashenok/2013} and white dwarfs \cite{das/2016}.

Anyhow, in \cite{harko/2011}, some flaws in the $f(R)$ gravity applicability were collected. It was brought to the scientific community attention the fact that the vast majority of the proposed $f(R)$ models are ruled out by Solar System regime tests \cite{erickcek/2006,chiba/2007,capozziello/2007}. Galactic scale tests also cannot support the $f(R)$ theory. Rotation curves of spiral galaxies were constructed in $f(R)$ gravity, but the results did not favor such a theory \cite{dolgov/2003,chiba/2003,olmo/2005}. Anyhow, it is worth mentioning that the first viable cosmological models of dark energy (but not dark matter) in f(R) theory were independently constructed in \cite{sawiki/2007,Appleby/2007,Starobinsky/2007}, all all of them avoiding the Dolgov-Kawasaki instability.

Motivated by these important issues, still in \cite{harko/2011}, it was proposed a generalization of the $f(R)$ theory, by including in the gravitational action, besides the general term in $R$, a general term in $T$, the trace of the energy-momentum tensor, yielding the $f(R,T)$ theory of gravity. The $T-$dependence of the theory can be due to the consideration of quantum effects, generating a potential path to a quantum theory of gravity \cite{xu/2016}. It can also be related to the existence of imperfect fluids in the universe \cite{harko/2011}.

It has been shown that such theories are capable of well describing the Solar System regime \cite{shabani/2014}. The dark matter galactic effects can be understood as due to the field equations extra terms predicted by the theory \cite{zaregonbadi/2016}. It was also shown that $f(R,T)$ gravity can give a considerable contribution to gravitational lensing \cite{alhamzawi/2016} and a deviation to the usual geodesic equation \cite{baffou/2017,ronaldo2018}. Although the cosmological viability of non-conservative $f(R,T)$ gravity models has been challenged recently, the same cannot be said for conservative models \cite{velten/2017}. Anyhow, the following works \cite{Jamil2012,Shabani2014,Baffou2014,Baffou2015,Sun2016} put $f(R,T)$ gravity non-conservative models in test in the observational cosmology aspect and the results are in favor of the theory. In \cite{Jamil2012}, the authors have shown that a dust fluid in $f(R,T)$ gravity reproduces $\Lambda$CDM cosmology. The obtained Hubble parameter shows good agreement with baryon acoustic oscillations. In \cite{Shabani2014}, it was shown that the functional form $f(R,T)=R+\alpha^{R-n}+\sqrt{-T}$, with constant $\alpha$ and $n$, yields consistent quantities when compared to observational data, for $n=-0.9$. We should also remark here that E.H. Baffou et al. in \cite{Baffou2014} have shown that the cosmological solutions of $f(R,T)$ gravity are consistent with the observational data for low and high redshift regimes. 

Furthermore, in \cite{Baffou2015}, the authors have obtained, from a non-standard Hubble parameter as a function of redshift, $H(z)$, acceptable models, also consistent with observational data. In \cite{Sun2016}, the authors have numerically simulated the plots of redshift versus distance modulus for different $f(R,T)$ models and obtained good fitting curve compared with astronomical observational data.

Moreover, the $f(R,T)$ cosmology evades the dark energy problem, by describing the cosmic acceleration as also due to the extra terms in $T$ in the field equations of the model \cite{mrc/2016,ms/2016,ms/2017,mcr/2018,singh/2016,kumar/2015}, instead of being due to the presence of a cosmological constant.

As we shall quantitatively visit later, an important consequence of the $f(R,T)$ field equations dependence on extra material (instead of geometrical) terms is the non-vanishing of the covariant derivative of the matter energy-momentum tensor, that is,  $\nabla_\mu T^{\mu\nu}\neq0$ \cite{harko/2011,barrientos/2014}. 

Cosmologically speaking, the non-conservation of the energy-momentum tensor is interpreted as due to creation (or destruction) of matter throughout the universe evolution \cite{singh/2016,kumar/2015}. This subject was deeply investigated from a thermodynamical perspective of the $f(R,T)$ gravity theory in \cite{harko/2014}. The same kind of physical property can be appreciated in another non-conservative energy-momentum theories, such as those presented in \cite{harko/2010,harko/2013}.

Very recently, the cause of the cosmic acceleration itself was proposed to be related with terms that do not conserve the energy-momentum tensor \cite{shabani/2017,josset/2017}.

In an astrophysical level, say, in the construction of the Tolman-Oppenheimer-Volkoff (TOV) equation \cite{tolman/1939,oppenheimer/1939}, \\$\nabla_\mu T^{\mu\nu}\neq0$ cannot be interpreted in the same way as in cosmology. Although the creation of matter in the universal scale mentioned above is a consequence of processes occurring in a quantum scale, the creation or destruction of matter particles shall not occur in a static analysis, such as the TOV equation.

Therefore one is left with the following question: is it possible to physically interpret by other means the non-conservation of the energy-momentum tensor in the hydrostatic equilibrium of a compact star? 

It has been shown that in order to construct the TOV equation in the $f(R,T)$ gravity, one needs the following non-conservative equation \cite{mam/2016,carvalho/2017}

\begin{equation}\label{i1}
	p'+(\rho+p)\frac{\phi'}{2}=-\frac{\lambda}{8\pi+2\lambda}(p'-\rho').
\end{equation}
In (\ref{i1}), $p$ and $\rho$ are the pressure and matter-energy density of the star, respectively, $\phi$ is a spherically symmetric metric potential, $'\equiv d/dr$ and it was taken $f(R,T)=R+2\lambda T$, with $\lambda$ a constant. However, as discussed above, the non-conservation of the energy-momentum tensor in the TOV equation construction cannot be interpreted as due to creation of matter. Therefore, Eq.(\ref{i1}) still lacks a somehow rigorous physical interpretation in the astrophysics context. In order to evade this problem, we will, for the first time in the literature, impose the conservation of the energy-momentum tensor in $f(R,T)$ gravity for the hydrostatic equilibrium of compact stars, and apply to neutron stars. That is the main purpose of the present article.

The article is organized as follows: in Section \ref{sec:frt}, we describe some important mathematical and physical properties of the $f(R,T)$ gravity. In Section \ref{sec:cfrt} we present a specific functional form for the function $h(T)$ in $f(R,T)=R+h(T)$ which conserves the energy-momentum tensor. We discuss some particular cases of this function. Once such a model is constructed, we derive and solve a new TOV equation in Section \ref{sec:tov}. We highlight and discuss our results in Section \ref{sec:con}.

\section{The $f(R,T)$ gravity}
\label{sec:frt}
Proposed as a generalization of the $f(R)$ theory, the $f(R,T)$ gravity has as its starting point the following action \cite{harko/2011}

\begin{equation}\label{frt1}
	\mathcal{S}=\int\left[\frac{f(R,T)}{16\pi}+\mathcal{L}_m\right]\sqrt{-g}d^{4}x,
\end{equation}
with $f(R,T)$ being the general function of $R$ and $T$, $\mathcal{L}_m$ is the matter lagrangian density, $g$ the determinant of the metric $g_{\mu\nu}$ and we are assuming natural units.

When varying this action with respect to $g_{\mu\nu}$, one obtains the following field equations

\begin{equation}\label{frt2}
	G_{\mu\nu}=8\pi T_{\mu\nu}+\frac{1}{2}h(T)g_{\mu\nu}+h_T(T)(T_{\mu\nu}-\mathcal{L}_mg_{\mu\nu}),
\end{equation}
in which $G_{\mu\nu}$ is the Einstein tensor, $T_{\mu\nu}$ is the energy-momentum tensor and we have already considered $f(R,T)=R+h(T)$, with $h(T)$ being a function of $T$ only, so that one recovers General Relativity in the regime $h(T)=0$. Moreover, $h_T(T)\equiv dh(T)/dT$.

The covariant derivative of $T_{\mu\nu}$ in (\ref{frt2}) reads

\begin{equation}\label{frt3}
	%\begin{align}
	\begin{split}
	&\nabla^{\mu}T_{\mu\nu}=\frac{h_T(T)}{8\pi+h_T(T)}[(\mathcal{L}_mg_{\mu\nu}-T_{\mu\nu})\nabla^{\mu}\ln h_T(T)+\\ & \nabla^{\mu}(\mathcal{L}_m-\frac{1}{2}T)g_{\mu\nu}].
	\end{split}
	%\end{align}
\end{equation}

\section{A conservative version of the $f(R,T)$ gravity}
\label{sec:cfrt}
It is natural to think that a certain form for $h(T)$ (besides the trivial one) in Eq.(\ref{frt3}) above conserves the energy-momentum tensor, i.e., yields $\nabla^{\mu}T_{\mu\nu}=0$. In order to find this form we can force the {\it lhs} of Eq.(\ref{frt3}) to be $0$ and search for the referred solution for $h(T)$.

We choose the matter lagrangian density to be $\mathcal{L}_m=\rho$ in order to have the energy-momentum tensor describing a perfect fluid and we also define the metric for a spherically symmetric object as

\begin{equation}\label{metric}
	ds^2=e^{\phi} dt^2-e^{\psi}dr^2-r^2d\theta^2-r^2\sin^2\theta d\phi^2,
\end{equation}
with $\psi=\psi(r)$ being a metric potential that depends on $r$ only.

From the assumptions above and taking $\nu=1$ in Eq.\ref{frt3}, we have

\begin{equation}\label{eqconv}
	(\rho+p)(\ln h_T)'+\frac{1}{2}(\rho+3p)'=0,
\end{equation}
where a prime indicates radial derivative and we have taken into account that $T=\rho -3p$ for a perfect fluid, which is being assumed. 

From now on, we need to define an equation of state (EoS) to be used in order to find the functional $h(T)$ that will satisfy \eqref{eqconv}. The EoS to be employed here is a polytropic relation between pressure and energy density. Such an EoS was used in the literature, by Tooper, to model the EoS of neutron stars \cite{tooper/1964}.

One of the most common EoS used in the literature to describe neutron stars is the polytropic one, where the relation between pressure and energy density assumes the form $p=k\rho^{\Gamma}$, where $k$ is a constant and $\Gamma$ the adiabatic index.
It can well represent the neutron stars we are concerned here, as we see in 
\cite{mam/2016,ray/2003}, and from this choice we can find a function $h(T)$ that conserves the energy-momentum tensor. 

In fact, the advantage of using such an equation of state (EoS), which leads to the maximum masses of neutron stars to be $\sim1.4M_\odot$, is that any increasing in this limit is due to the gravity theory itself. 

It is important to stress that we are using a polytropic EoS in the energy density that is really different from a         opic one in the particle number density or mass density.
It is know that the last case, usually used in white dwarf calculations or Newtonian stars, only by applying first law of thermodynamics, it always produce a linear relation between pressure and energy density ($\Gamma =1 $)  \cite{lai/2009}, that is not a good approximation for neutron stars, where the sound velocity is not constant when the energy density changes inside the star. 

From \eqref{eqconv}, and assuming the polytropic EoS, one can obtain
\begin{equation}
h_T(\rho)=\alpha\left(\rho+k \rho^\Gamma\right)^{\frac{\left(1-3\Gamma\right)}{2(\Gamma-1)}} \rho^{\frac{\Gamma}{(\Gamma-1)}} .
\end{equation}
and, consequently,
%\begin{equation}
%  h(\rho)=\int \alpha\left(1-3k\Gamma\rho^{\Gamma-1}\right)\left(\rho+k \rho^\Gamma\right)^{\frac{\left(1-3\Gamma\right)}{2(\Gamma-1)}} \rho^{\frac{\Gamma}{(\Gamma-1)}} d\rho,
%\end{equation}

\begin{scriptsize}
\begin{equation}\label{funct}
\begin{split}
h(\rho) = \alpha \frac{1}{3\Gamma-2}2\rho^{\frac{\Gamma}{\Gamma-1}}\left(\frac{\rho^{1-\Gamma}+k}{k}\right)^{\frac{1-3\Gamma}{2(1-\Gamma)}} \left(\rho + k\rho^\Gamma\right)^{\frac{1-3\Gamma}{2(\Gamma-1)}} \times \\ \left[3(3\Gamma-2)k\rho^\Gamma ~_2F_1\left(\frac{1-3\Gamma}{2(1-\Gamma)},\frac{\Gamma}{2(\Gamma -1)};\frac{\Gamma}{2(\Gamma -1)}+1;-\frac{\rho^{1-\Gamma}}{k}\right)\right. \\ \left.-\rho ~_2F_1\left(\frac{1-3\Gamma}{2(1-\Gamma)},\frac{2-3\Gamma}{2(1-\Gamma)};\frac{4-5\Gamma}{2(1-\Gamma)};-\frac{\rho^{1-\Gamma}}{k}\right)\right],
\end{split}
\end{equation}
\end{scriptsize}
where $_2F_1$ represents the Gauss hyper-geometric function. It is worth to cite that as $T(\rho)$ one can construct the functional $h(T()$ by performing the numerical calculation of \eqref{funct}.

Following the recent works  to describe neutron stars \cite{mam/2016,ray/2003}, we can consider $\Gamma=5/3$, and find the functional $h$
\begin{equation}
h_T(\rho)=\alpha \frac{\rho^{5/2}}{\left(\rho+k\rho^{5/3}\right)^3},
\end{equation}

\begin{small}
\begin{equation}
\begin{split}
h(\rho) = \frac{3\alpha}{64} \left[ \frac{5\sqrt{2}\log \left(\sqrt{k}\sqrt[3]{\rho} + \sqrt{2}\sqrt[4]{k}\sqrt[6]{\rho}+1\right)}{k^{3/4}}  \right. \\ - \frac{5\sqrt{2}\log \left(\sqrt{k}\sqrt[3]{\rho} - \sqrt{2}\sqrt[4]{k}\sqrt[6]{\rho}+1\right)}{k^{3/4}}+ \frac{10\sqrt{2}\tan^{-1}\left(1-\sqrt{2}\sqrt[4]{k}\sqrt[6]{\rho}\right)}{k^{3/4}}  \\ \left.- \frac{10\sqrt{2}\tan^{-1}\left(1+\sqrt{2}\sqrt[4]{k}\sqrt[6]{\rho}\right)}{k^{3/4}} -\frac{40\rho}{k\rho^{2/3}+1} + \frac{96\sqrt{\rho}}{(k\rho^{2/3}+1)^2} \right].
\end{split}
\end{equation}
\end{small}

\section{The TOV equation and its solutions from a conservative $f(R,T)$ gravity}\label{sec:tov}

By using the metric defined in (\ref{metric}) we obtain the $tt$ and $rr$ components of the field equations as

\begin{subequations}\label{eq:camp:3}
	\begin{align}
		\begin{split}
			&\frac{e^{-\psi}}{r^{2}}(e^{\psi}+\psi'r-1)=8\pi \rho + \frac{1}{2}h(\rho),
		\end{split}
		\label{eq:camp:tt}
		\\
		\begin{split}
			&\frac{e^{-\psi}}{r^2}\left(1-e^{\psi}+\phi'r\right)=8\pi p - \frac{1}{2}h(\rho) -h_{T}(p-\rho).
		\end{split}
		\label{eq:camp:rr}
	\end{align}
\end{subequations}

We introduce now the quantity $m(r)$ which depends on the radial coordinate only, such that
\begin{equation}
	e^{-\psi}=1-\frac{2m}{r},
\end{equation}
and replacing it into (\ref{eq:camp:tt}) we get
\begin{equation}\label{masseq}
	\frac{m'}{r^2}=4\pi \rho + \frac{1}{4}h(\rho).
\end{equation}

Let us recall that from the conservation of the energy-momentum tensor we have:

\begin{equation}
	\nabla^{\mu}T_{\mu\nu}=-p' -(\rho+p)\frac{\phi'}{2}=0.
\end{equation}

By isolating $\phi'$ in (\ref{eq:camp:rr}), one is able to derive the modified TOV equation as follows

\begin{equation}\label{newtov}
	p'=-(\rho+p)\dfrac{\left\{\frac{m}{r^2}+ \left[4\pi p - \frac{1}{4}h(\rho) -\frac{1}{2}h_{T}(p-\rho)\right]r\right\}}{1-\frac{2m}{r}},
\end{equation}
if $\alpha=0$ we canceling out any contribution from the trace of the energy-momentum tensor in the field equations. 

The Equations (\ref{masseq}) and (\ref{newtov}) can be solved numerically by using a Runge-Kutta method. The boundary conditions at the center of the star are as follows: $p(0)=p_c$, $\rho(0)=\rho_c$ and $m(0)=0$, with $p_c$ and $\rho_c$ being the central pressure and central energy density. For $r=R$, where the pressure and energy density of the star vanish, the enclosed mass $m(R)=M$ will represent the total mass of the star and $R$ its total radius. By using different values of central energy density, one is able to construct the mass-radius relation as well as other relations that we further derive in this work.

Figure \ref{massxradius} below shows the behaviour of the total mass with total radius of the neutron star, where several values of $\alpha$ were used and $\Gamma=5/3$ in reference to the work of R.F. Tooper \cite{tooper/1964}, and also in \cite{mam/2016,ray/2003}. It is worth to quote that $\alpha=0$ corresponds to results found within General Relativity theory. 

\begin{figure}[h!]
	\begin{center}
		\includegraphics[width=1.0\linewidth]{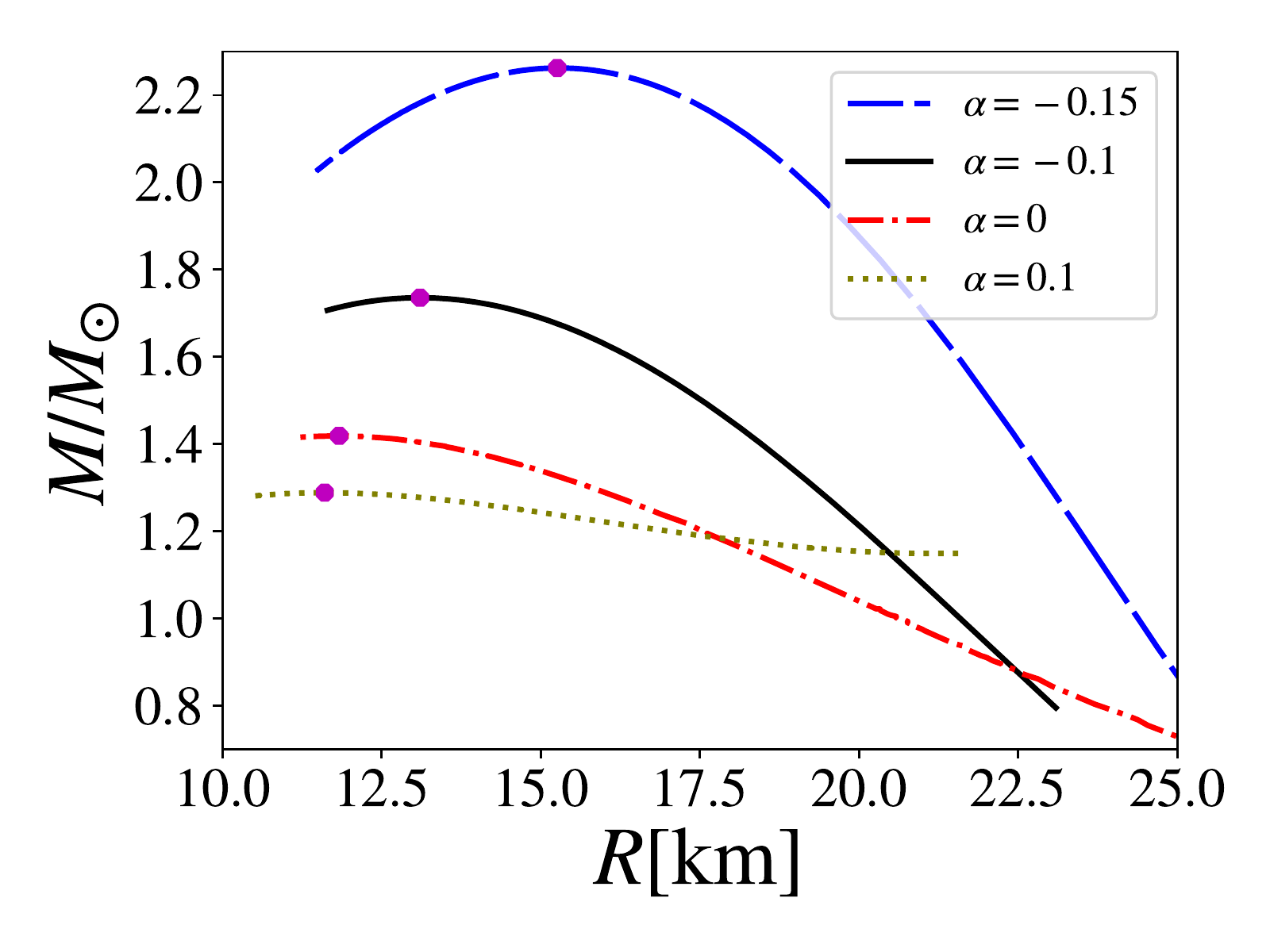}
		\caption{Mass-radius neutron star relation for the conservative model of $f(R,T)$ gravity. Several values of $\alpha$ were employed and $\Gamma=5/3$.}
		\label{massxradius}
	\end{center}
\end{figure}

From Fig.\ref{massxradius} we can note that for larger positive $\alpha$, less massive and smaller stars are found. Such a behaviour can be understood as a strong gravity regime effect, as the case of GR in comparison with Newtonian gravitation \cite{carvalho/2018}. On the other hand, for the cases where $\alpha<0$ we observe an increasing in the total mass and total radius of the neutron star when $\vert\alpha\vert$ increases.

\begin{figure}[h!]
	\begin{center}
		\includegraphics[width=1.0\linewidth]{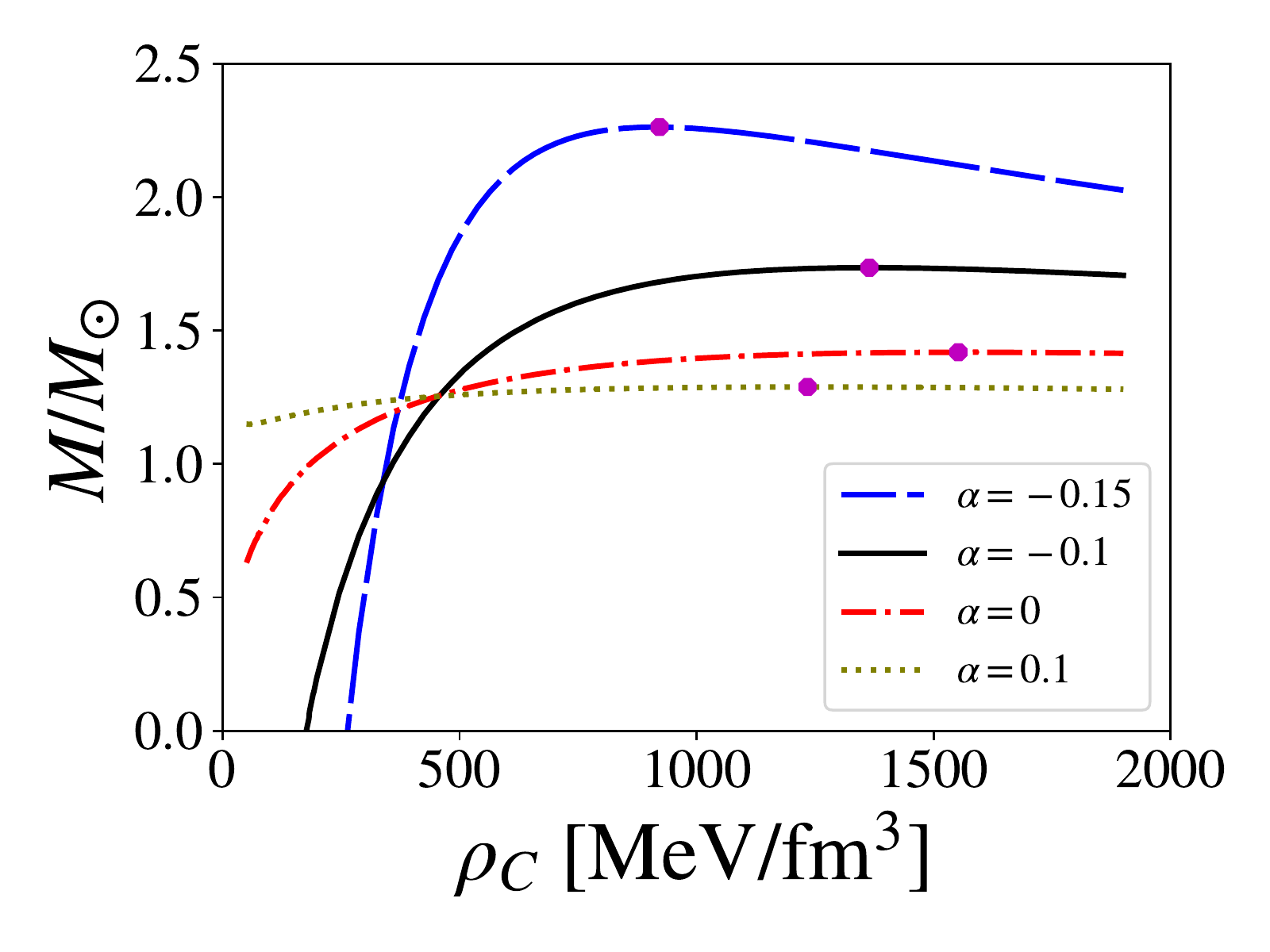}
		\caption{Total mass versus central energy density of neutron stars ($\Gamma=5/3$) for the conservative model of $f(R,T)$ gravity using several values of $\alpha$. The maximum mass points mark the onset of instability.}
		\label{massxdensity2}
	\end{center}
\end{figure}

In Fig.\ref{massxdensity2} below we show the neutron star mass against central energy density for the conserved model of $f(R,T)$ gravity, where several values of $\alpha$ were employed. For all values of $\alpha$ we observe that the mass initially increases with central density until it attains a maximum value. After that point, the mass decreases with the increasing of central density. 

From the regular criterion of stability, $\partial M/\partial\rho_c>0$, we conclude that the maximum mass points mark the onset of instability in the curves of Fig.\ref{massxdensity2}. In addition, the value of $\lambda=-0.15$ produces a maximum mass stable neutron star of $M\approx 2.2M_{\odot}$.

\section{Discussion}\label{sec:con}

Alternative gravity theories offer an important possibility for solving or evading some issues one faces when treating General Relativity as the background theory of gravity. On this regard, one could check Refs.\cite{capozziello/2011,capozziello/2008,bull/2016}.

In the present paper we have worked with a particular alternative gravity, named $f(R,T)$ theory. The $T$ dependence of the theory is due to the consideration of quantum effects or to the possible existence of imperfect fluids in the universe. Among other features, the $f(R,T)$ gravity predicts a non-conservation of the energy-momentum tensor. Anyhow, one can find a few $f(R,T)$ gravity references in which the energy-momentum tensor is conserved through the imposition of some mechanism or particular approach.

F.G. Alvarenga et al. in the Reference \cite{alvarenga/2013} have imposed $\nabla^{\mu}T_{\mu\nu}=0$ and derived the resulting dependence on $T$ for the function that conserves the $f(R,T)$ energy-momentum tensor in cosmology, for a Friedmann-Lema\^itre-Robertson-Walker metric, as $f(T)=\alpha T^{\frac{1+3c_s^2}{2(1+c_s^2)}}$, for which $\alpha$ is an integration constant and $c_s=\sqrt{\partial p/\partial\rho}$ is the speed of sound. S. Chakraborty has shown that when $f(R,T)=A(R)+B(T)$, with $A$ and $B$ being functions of their arguments and $A(R)\neq R$, the function $B(T)=B_0T^{\frac{1}{1+\omega}}$, with $B_0$ an integration constant, conserves the $f(R,T)$ energy-momentum tensor \cite{chakraborty/2013}.

In \cite{mcr/2018}, it was proposed a different mechanism to conserve the $f(R,T)$ gravity energy-momentum tensor for a particular function, named $f(R,T)=R+2\chi T$. It was shown that what makes $\nabla^{\mu}T_{\mu\nu}\neq0$ in the $f(R,T)$ gravity is the indistinct presence of $\tilde{T}_{\mu\nu}$ defined as

\begin{equation}\label{ef6}
	\tilde{T}_{\mu\nu}\equiv\frac{\chi}{8\pi}[2(T_{\mu\nu}+pg_{\mu\nu})+Tg_{\mu\nu}],
\end{equation}
such that the theory may be described by an effective energy-momentum tensor given by $T_{\mu\nu}^{\texttt{eff}}=T_{\mu\nu}+\tilde{T}_{\mu\nu}$. By using simple mathematical properties, it was shown that both $\nabla^{\mu}T_{\mu\nu}=0$ and $\nabla^{\mu}\tilde{T}_{\mu\nu}=0$ can be simultaneously satisfied, with the latter indicating the existence of stiff matter fluid permeating the universe together with the ordinary energy-momentum tensor fluid.

In particular, in this work, we derived the functional $h(T)$ that conserves the energy momentum-tensor by considering the specific model $f(R,T)=R+h(T)$ and a polytropic EoS in the energy density $\rho$ given by $p=k\rho^{\Gamma}$, with $k$ being the EoS parameter. We have derived the hydrostatic equilibrium equations for the conservative model of  $f(R,T)$ gravity here obtained, showing that it presents new terms in comparison to the standard TOV equation. 

Regarding phenomenology of neutron stars ($\Gamma=5/3$), we showed that their structure suffers significant deviations from general relativistic results and depends on the choice of the value of the free parameter $\alpha$. In particular, for negative values of $\alpha$ we found larger and more massive stars with the increasing of $|\alpha|$. On the other hand for positive values of $\alpha$ we obtained smaller and less massive stars according to the increasing of $\alpha$. These features can be appreciated in Fig.\ref{massxradius}.

In Fig.\ref{massxdensity2} we have plotted the neutron star mass against its central density. The figure refers to our conservative model and this presents a sensitive contribution to the increasing of mass for $\alpha<0$.

Furthermore, in Fig.\ref{massxdensity2}, together with the regular criterion of stability, indicates a lower limit for $\alpha$ in the present model, which reads $-0.15$.

In summary, we have presented for the first time in the literature the hydrostatic equilibrium configurations of neutron stars for the $f(R,T)$ gravity with conservation of the energy-momentum tensor. We have shown that by imposing $\nabla^{\mu}T_{\mu\nu}=0$ in such a theory, a unique function of $T$ in $f(R,T)$ is obtain for each EoS one may use. Particularly, in this work we choose a quite general one, $p=k\rho^{\Gamma}$. Several works have been done before concerning compact objects in $f(R,T)$  without imposing the energy-momentum conservation. In these cases, for a specific EoS choice it was possible to use any form for the $h(T)$ function. In our approach, this is not true anymore, which is a manifestation of the existence of a strong connection between the conserved $f(R,T)$ gravity and the star matter constitution. Thus, when we have a complete knowledge of the equation of state of cold super dense matter, we will be able to find an unique form for $h(t)$ in the conserved $f(R,T)$ gravity and construct a unified theory for astrophysics and cosmology.  

We also obtained in our approach, for the particular case of $\alpha<0$, a mass $\times$ radius diagram in which high neutron star masses, with large radius, can be attained. As in the work done before of neutron stars in a non conservative $f(R,T)$ gravity, the high neutron star masses are obtained with lower central energy densities $\rho_c$ \cite{mam/2016}.  In view that recent astrophysical observations of massive pulsars still await convincing explanations, a mechanism able to provide such an increasing on the mass is valuable.  Also, further searches for novel conservative functional forms for the $f(R,T)$ function can be attained in near future, from different equations of state.

Finally, we would like to stress that we cannot constrained the cold dense matter equation of state by astronomical observations of very massive pulsars, since it is possible to obtain high neutron stars masses with an EoS that in the normal General Relativity theory will produce only a maximum mass of 1.4 $M_{\odot}$, as it is the case of the EoS used here. The only way to know the high density behaviour of nuclear matter is by nuclear physics experiments in the laboratory, since the maximum neutron star mass depends also on the gravity theory used, as we have shown here.

\bigskip

\section{Acknowledgments}
	We would like to thank the Referee for his/her constructive comments. The authors thank the financial support of S\~ao Paulo Research Foundation (FAPESP) under the thematic project 2013/26258-4. PHRSM would like to thank also (FAPESP), grant 2015/08476-0. GAC thanks CAPES for financial support under the process 88881.188302/2018-01.

\end{document}